\documentclass[fleqn,12pt,twoside]{article}
\usepackage{amssymb}
\usepackage{espcrc1}
\usepackage{graphicx}
\usepackage{wrapfig}

\newcommand{\npart}{\mbox{$N_{part}$}}
\newcommand{\nbin}{\mbox{$N_{bin}$}}
\newcommand{\dnchdeta}{\mbox{$dN_{\mathrm{ch}}/d\eta$}}
\newcommand{\sqrts}{\mbox{$\sqrt{s}$}}
\newcommand{\sqrtsNN}{\mbox{$\sqrt{s_{_{\mathrm{NN}}}}$}}

\newcommand{\ppbar}{\mbox{$p\bar{p}$}}
\newcommand{\ee}{\mbox{e$^+$e$^-$}}
\newcommand{\pp}{\mbox{$pp$}}

\newcommand{\meanpt}{\mbox{$\langle p_T \rangle$}}
\newcommand{\pt}{\mbox{$p_T$}}
\newcommand{\Et}{\mbox{$E_T$}}
\newcommand{\meanEt}{\mbox{$\langle E_T \rangle$}}
\newcommand{\meanNch}{\mbox{$\langle N_{\mathrm{ch}} \rangle$}}

\newcommand{\gevc}{\mbox{${\mathrm{GeV/}}c$}}
\newcommand{\mevc}{\mbox{${\mathrm{MeV/}}c$}}

\title{Experimental summary on global observables, hadron spectra and ratios}

\author{Thomas S. Ullrich \address{
        Brookhaven National Laboratory,
        Upton New York 11973-5000, USA}}%

\begin{document}

\maketitle

\section{Introduction}
\label{sec:intro}

In this article I summarize results on global observables, hadron
spectra, and ratios of integrated hadron yields as presented at the
Quark Matter 2002 Conference.  I also attempt to put these results
into context and convince the reader that an evolving coherent picture
begins to form, shedding light on the state of matter created in
relativistic heavy ions collisions. However, at this conference we
have been presented with such a wealth of new data on hadronic signals
that no summary can give proper credit to everyone.  A definitive
summary of all the results on hadron spectra, ratios, and yields and
their interpretation would certainly require a much longer paper. I
therefore limit myself to a compilation of results that I consider to
be the most interesting, and refer the reader to the large number of
excellent papers given at the conference for further details.

It was obvious that this conference was dominated by the latest
findings from the RHIC experiments. Despite the short time span
between the last run and the Conference, spectra of many particle
species were reported, both over a wide range in \pt\ and, for the
first time, also systematically over a broad range of rapidity. The
longer RHIC run in 2001/2002, delivering for the first time the design
energy of $\sqrtsNN = 200$ GeV, allowed the experiments to accumulate
more statistics and, together with the progress in the understanding
of the detectors, resulted in higher precision measurements than
possible at the lower energy ($\sqrtsNN = 130$ GeV) studied in the
previous run in 2000/2001.

Although the SPS heavy ion program at CERN has been almost completed,
with all experiments delivering the physics they were built for, work
on the analysis of data, their systematic comparison and
interpretation continues.  In order to understand the physics behind
the state of matter produced in relativistic heavy ion collisions one
must, more than everything else, understand the excitation functions
of all observables involved. Only if we are able to describe the
evolution and behavior of the systems as we increase the energy --
from AGS, through SPS to RHIC -- will we gain insight into the rich
variety of physics we are facing.

In this article I start with a summary of global observables presented
at this conference followed by a discussion of particle ratios and
chemical freeze-out conditions.  I then address the question of
boost-invariance at RHIC and conclude with a discussion of transverse
momentum spectra and kinetic freeze-out parameters.

\section{Global Observables}
\label{sec:global}

One of the earliest probes suggested for QGP formation involves a
study of the global parameters of the events, e.g.~the energy
deposition, multiplicity, and the average transverse momentum of the
emitted particles, as a function of center-of-mass energy \sqrtsNN,
mass number $A$, and centrality of the collision.  For example, by
studying the multiplicity of the produced particles one might estimate
theoretically the entropy produced in the collision. Sudden changes in
behavior with varying centrality or $A$ would be indicators of a phase
transition. So far, however, no such anomalous changes have been
observed, at either the AGS, SPS, or RHIC. All results on global
observables shown at this conference indicate a rather smooth
evolution in centrality and \sqrtsNN. This, of course, does not
necessarily imply the absence of a phase transition, but might be
rather an indication of either the insensitivity of these observables
to the early phase of the collision and/or might suggest a second
order phase transition (or a cross-over).

\begin{wrapfigure}{l}{0.5\textwidth}
    \vspace{-7mm}
    \includegraphics[width=0.48\textwidth]{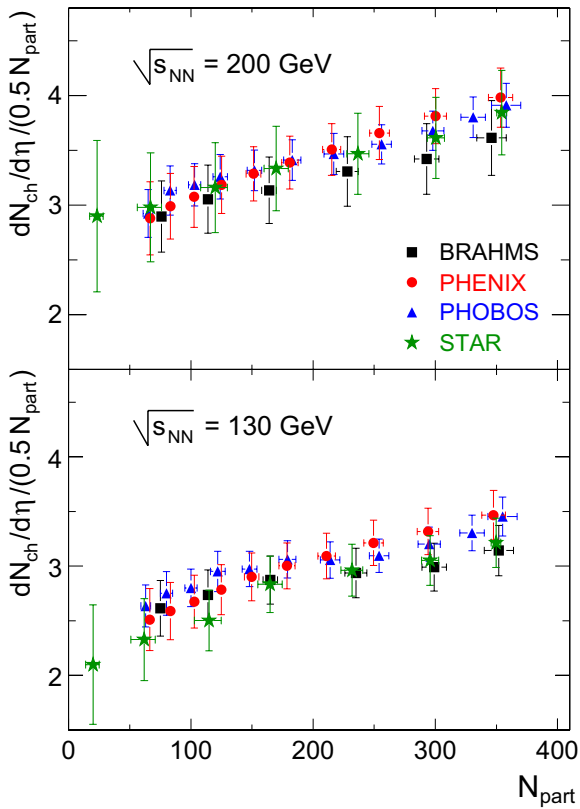}
    \vspace{-3mm}
        \caption{\footnotesize Charged particles per participant pair as a
            function of number of participants for $\sqrtsNN = 130$ and
            200 GeV for all four RHIC experiments (from
            \cite{sasha1,calderon}).  PHENIX and STAR data are
            preliminary.}
        \label{fig:nchRHIC}
    \vspace{-7mm}
\end{wrapfigure}
With the commencement of the RHIC program the question of
multiparticle production in nuclear collisions became more complex due
to the poorly understood role of perturbative QCD (hard processes).

The study of charged particle multiplicity as a function of the number
of participating nucleons was conducted by all four RHIC experiments
and the results presented in Fig.~\ref{fig:nchRHIC} depict the high
quality and agreement between the experiments \cite{sasha1}. It should
be noted that, even more than in the determination of the multiplicity
and the required corrections for decays, absorption, and feed-down,
the difficulties of this analysis lie in the determination of \npart,
\textit{i.e.}~the extraction of the underlying collision geometry from
the data. This requires a detailed understanding of the trigger,
especially trigger efficiencies, contamination and the study of
possible auto-correlations.  All four RHIC experiments now use
Monte-Carlo Glauber model calculations similar to the ones implemented
in the Hijing model as compared to numerical calculations of the
nuclear overlap functions in the optical limit; both approaches
disagree slightly, with the latter having problems estimating the
total inelastic cross-sections.  For the ratio of multiplicities at
mid-rapidity between the two energies, $\sqrtsNN = 200$ GeV and 130 GeV,
the experiments report values of $1.14 \pm 0.05$ (PHOBOS \& BRAHMS),
$1.22 \pm 0.08$ (STAR), and $1.17 \pm 0.03$ (PHENIX) with no
indication of any significant centrality dependence
\cite{foot1} \cite{sasha1,baker,bearden,gene}.

\begin{wrapfigure}{l}{0.5\textwidth}
    \vspace{-2mm}
    \includegraphics[width=0.48\textwidth]{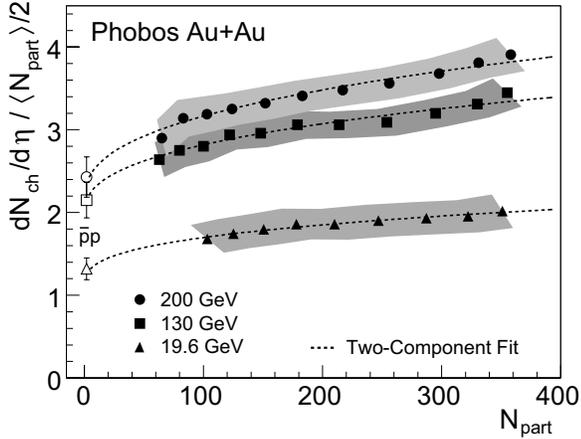}
    \vspace{-2mm}
    \caption{\footnotesize Charged particles per participant pair
        as a function of number of participants in $\sqrtsNN = 19.6,
        130$ and 200 GeV Au+Au collisions measured by PHOBOS
        (from \cite{baker}). The curves are two-component fits
        described in the text and in \cite{kn}.}
    \label{fig:PhobosCentDep}
    \vspace{-5mm}
\end{wrapfigure}
The interpretation of the scaling of the multiplicity at mid-rapidity
as a function of \npart\ appears still ambiguous. A simple model by
Kharzeev and Nardi (KN) explains the dependence in a two-component
approach differentiating between soft processes scaling with \npart\ 
and hard processes scaling with \nbin\ \cite{kn}. When fit to the data
as shown in Fig.~\ref{fig:PhobosCentDep} by the PHOBOS collaboration
the model allows one to extract the fraction of particles produced
from hard processes and one obtains values of 36\% for $\sqrtsNN =
130$ GeV, and 45\% for 200 GeV, respectively \cite{baker,prc-phobos}.
A second class of calculations is based on parton saturation
\cite{eskola,levin}; since the parton densities in the initial stage
of the collision can be related to the density in the final state a
parametrized dependence of the saturation scale $Q_s$ on \sqrts\ and
impact parameter allows one to predict \dnchdeta.  However, the
predictions from these and related models have been found to be almost
indistinguishable when applied to RHIC data, especially because of the
large experimental uncertainties in the calculation of \npart\ for
very peripheral collisions, the region where the differences between
the various models become more apparent. An exception are models based
on final state saturation who significantly overpredict the yield at
low \npart.  The systematic uncertainties in \npart\ for peripheral
events can only be reduced when data from collisions of light ions,
A$<100$, become available.

\begin{wrapfigure}{r}{0.5\textwidth}
    \vspace{-8mm}
    \includegraphics[width=0.48\textwidth]{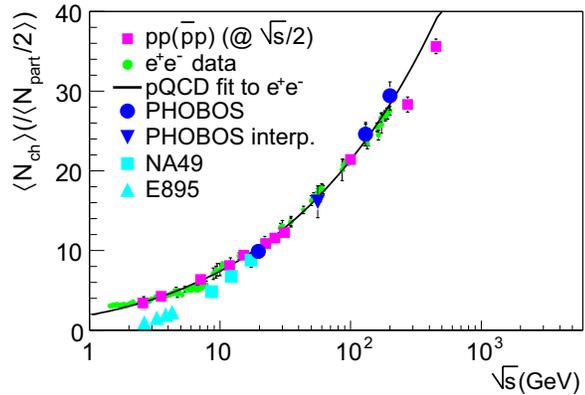}
    \vspace{-2mm}
    \caption{\footnotesize Total charged multiplicity per participant
        pair as a function of \sqrts\ for \ee, \pp\ (\ppbar), and
        AA collisions. The \pp\ (\ppbar) data is plotted versus
        $\sqrts_{\mathrm{eff}} = \sqrts/2$ (from \cite{steinberg}).}
    \label{fig:totaleeAApps}
    \vspace{-5mm}
\end{wrapfigure}
A very surprising finding in the context of particle production was
presented by the PHOBOS collaboration at this conference
\cite{baker,steinberg}. The authors noted that the total charged
multiplicity per participant in elementary \ee\ and central AA
collisions is identical over a wide range of \sqrts. That alone is a
very remarkable fact but they also found that the same holds for the
\pp\ (\ppbar) collisions \emph{when} compared to the effective
center-of-mass energy $\sqrt{s_{\mathrm{eff}}}$, that is the nominal
energy minus the energy carried away by the leading protons. The
authors actually used $\sqrt{s_{\mathrm{eff}}} = \sqrts/2$ which was
verified by PYTHIA simulations.  This is depicted in
Fig.~\ref{fig:totaleeAApps}. Also shown is a perturbative QCD
calculation for the multiplicity in \ee\ fit to the data.

The agreement between the three fundamentally different collision
systems is remarkable, with the exception of AA collisions below
$\sqrtsNN = 20$ GeV and \pp\ collisions above 200 GeV. One might
speculate if this agreement is plainly accidental or if it possibly
points to a kind of universality in the production of
multihadronic final states in high energy collisions of \emph{any}
elementary particles over some range in \sqrts. It is even more
surprising that this should also apply for heavy ion collisions where
we have indications that the system thermalizes and evolves on a time
scale significantly larger than in elementary collisions and where the
majority of particles are formed late. In comparison, the hadron
production in \ee\ is mostly from hard gluon radiation (2-,3-and 4-jet
events), while in \pp\ one observes a mixture of soft and hard
processes.

\begin{wrapfigure}{r}{0.5\textwidth}
    \vspace{-7mm}
    \includegraphics[width=0.48\textwidth]{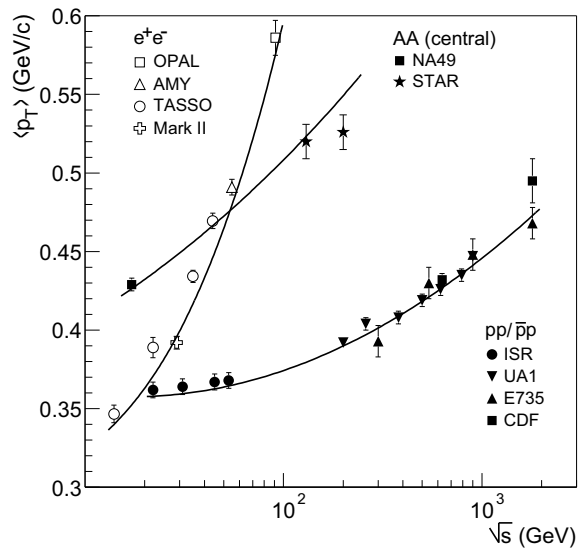}
    \vspace{-2mm}
        \caption{\footnotesize Mean transverse momentum versus \sqrts\ for
            \ee, \pp\ (\ppbar), and AA collisions (from \cite{gene,opal}).
            In \ee\ the \meanpt\ is measured with respect to the thrust axis.
            The curve through the \ee\ data points are JETSET predictions,
            the curve through the AA data (NA49 and STAR at 130 GeV)
            is from CGC predictions (see text), and that through the
            \pp\ (\ppbar) data is a phenomenological parameterization in
            log(\sqrts).}
        \label{fig:meanpt}
    \vspace{-5mm}
\end{wrapfigure}
To shed more light on this apparent similarity it is instructive to
look at the scaling of the average transverse momentum of the produced
particles in all three systems with \sqrts\ \cite{xzb}. Any
universality in the production mechanism should, to some degree, also
manifest itself in the evolution of \meanpt. However, this seems not
to hold as illustrated in Fig.~\ref{fig:meanpt} \cite{gene}.  The
figure shows that the \sqrts\ dependence of \meanpt\ in \ee\ is
considerably steeper than in \pp\ (\ppbar), mainly due to the absence
of soft processes in \ee\ collisions. This suggests that the agreement
in particle multiplicity might be accidental.  Still, the fact remains
that the total multiplicity per participant appears to be remarkably
similar for \ee, nucleon-nucleon, and nucleus-nucleus collisions and
further studies might help to gain insight into multihadron production
in high energy collisions.

Another interesting finding depicted in Fig.~\ref{fig:meanpt} is the
relatively small increase of \meanpt\ between $\sqrtsNN = 130$ and 200
GeV of only $\sim 1$\%, as pointed out by the STAR collaboration
\cite{gene}. Gluon saturation models and hydrodynamics predict a
considerably stronger dependence, usually $\meanpt^2 \propto
\dnchdeta$.  The solid curve is a prediction based on the
saturation model, constraint by \pp\ (\ppbar) results \cite{xzb}. The
200 GeV data point clearly falls below this prediction suggesting a
flattening of the \meanpt\ energy dependence at RHIC. The difference is
significant since the systematic errors for the 130 and 200 GeV data
points are correlated.  However, in order to prove or falsify any
model it is essential to perform further measurements at energies
between 20 and 200 GeV to study the scaling of \meanpt\ in greater
detail.

\begin{wrapfigure}{r}{0.5\textwidth}
    \vspace{-5mm}
    \includegraphics[width=0.48\textwidth]{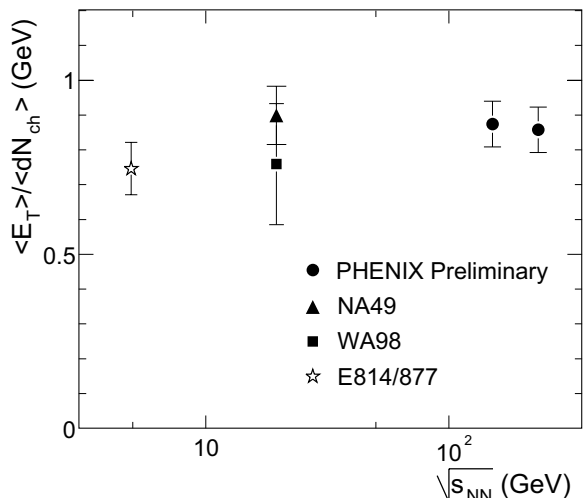}
    \vspace{-4mm}
        \caption{\footnotesize
            $dE_{T}/d\eta|_{\eta=0}/dN_{ch}/d\eta|_{\eta=0}$
            versus \sqrtsNN\
            for 5\% most central events at AGS, SPS,
            and RHIC (from \cite{sasha1}).}
        \label{fig:phenixEtS}
    \vspace{-7mm}
\end{wrapfigure}

Another important observable for characterizing the global properties
of bulk matter is the transverse energy \Et. This was studied in
detail by the PHENIX collaboration for $\sqrtsNN =130$ and 200 GeV
\cite{sasha1} in the mid-rapidity region. They find that $d\Et/d\eta$
and \dnchdeta\ increase with \npart\ in a very similar fashion
resulting in an almost constant ratio \meanEt/\meanNch $\sim 0.9$ GeV.
This holds for $\sqrtsNN = 130$ \emph{and} 200 GeV.  Even more surprising is
the fact that studies from Au+Au collisions at $\sqrtsNN = 4.8$ and
Pb+Pb collisions at 17.2 GeV yield very similar values, suggesting
that the increased energy put into the system results solely in an
increased particle production leaving the average energy per particle
almost constant. This is depicted in Fig.~\ref{fig:phenixEtS}. From
the measured $d\Et/d\eta$ for the 2\% most central Au+Au collisions at
200 GeV the authors estimated the Bjorken energy density to be
$\varepsilon_{\mathrm{BJ}} \approx 5.5$ GeV/fm$^3$, assuming a
conservative formation time of $\tau = 1$ fm/$c$. Similar studies at
SPS in Pb+Pb collisions at $\sqrtsNN = 17.2$ GeV give
$\varepsilon_{\mathrm{BJ}} \approx 3.2$ GeV/fm$^3$ \cite{na49Et}.
These values represent of course only a lower limit for the initial
energy density since the longitudinal expansion of the system reduces
the transverse energy considerably.  Recent lattice results on QCD
thermodynamics estimate the criticial energy density to be
$\varepsilon \approx 0.70 \pm 0.35$ GeV/fm$^3$ \cite{karsch}, a value
significantly surpassed already at SPS.

It is interesting to compare the scaling of \meanEt/\meanNch\ with
that of \meanpt\ depicted in Fig.~\ref{fig:meanpt}. While the small
increase in \meanpt\ between $\sqrtsNN = 130$ and 200 GeV is reflected by
the corresponding \meanEt/\meanNch\ remaining constant, the
considerably lower value of \meanpt\ at CERN/SPS energies appears to
be contradicted by the transverse energy per particle.
This, however, could be accounted for by the different particle
composition at SPS and RHIC, although it still needs to be verified in
quantitative studies.  It is intriguing to compare the universality
in \meanEt/\meanNch\ with an observation presented in a paper by
Cleymens and Redlich in 1998 \cite{cleymans} in which the authors
show that the chemical freeze-out parameters obtained at SPS, AGS, and
SIS all correspond to a unique value of $\sim 1$ GeV for the average
energy per hadron in the local rest frame of the system independent of
beam energy and mass number. From what we have learned so far this empirical
observation holds still at RHIC energies leading to a considerable
unification in the description of hadronic final states in high energy
nuclear collisions.

\section{Particle Ratios and Chemical Freeze-Out}
\label{sec:yields}

One of the most important issues in the physics of heavy ion
collisions is the question if, and if so at what stage, the produced
system thermalizes and to what extent a thermal description is
appropriate for the evolving system.  In order to discuss an
equation of state and a true order to any associated phase transitions,
we need to describe the system in terms of a few thermodynamic
properties.  The use of thermodynamic concepts to describe multi-particle
production has a long history beginning with Hagedorn in the early
1960's \cite{hagedorn}.  The concept of a \textit{temperature} applies,
strictly speaking, only to systems in at least local thermal
equilibrium. Thermalization is normally only thought to occur in the
transverse degrees of freedom as reflected in the Lorentz invariant
distributions of the particles. The measured hadron spectra contain two
pieces of information: \textit{(i)} their normalization,
\textit{i.e.}~their yields and ratios, provide the chemical
composition of the fireball at the chemical freeze-out point and
\textit{(ii)} their transverse momentum spectra which provide
information about thermalization of the momentum distributions and
collective flow. It is obvious that the observed single particle
spectra do not reflect earlier conditions, \textit{i.e.} the hot and
dense deconfined phase, where chemical and thermal equilibrium may
have been established, since rescattering erases most traces from the
dense phase. Only those effects which are accumulative during the expansion,
such as flow, remain.

\begin{figure}[t]
    \begin{center}
        \vspace{-5mm}
        \includegraphics[width=\textwidth]{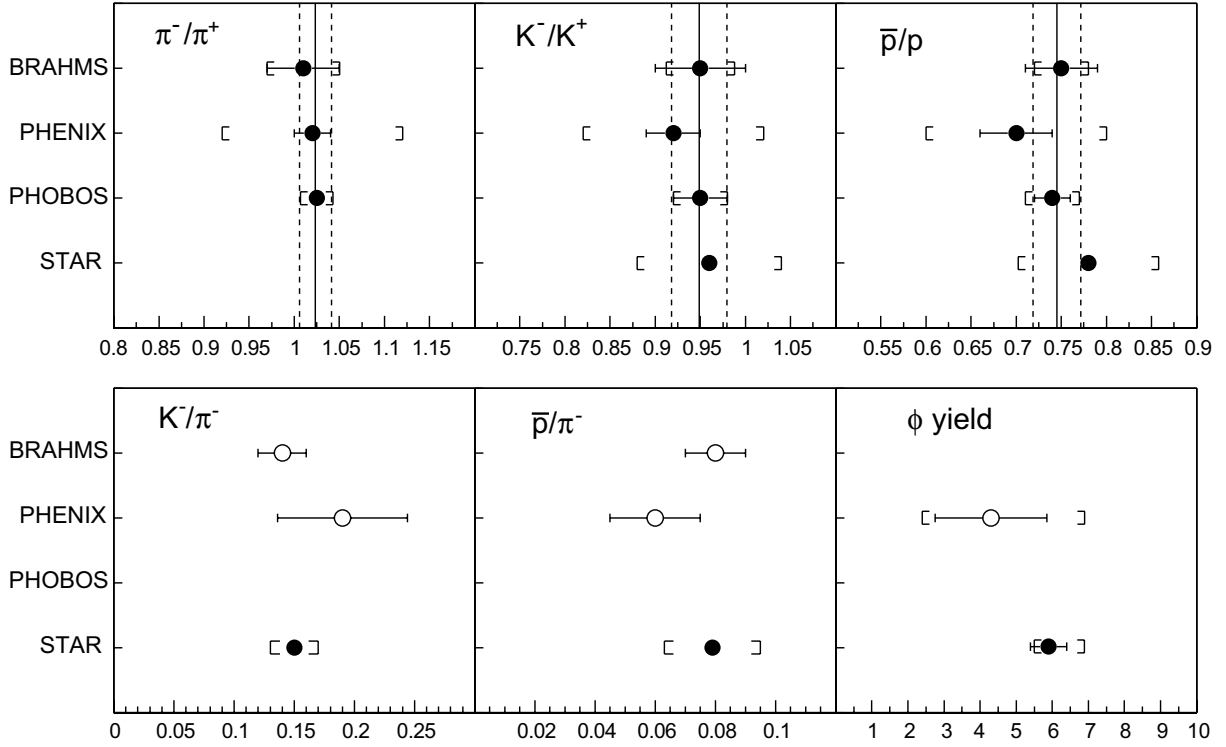}
        \vspace{-7mm}
        \caption{\footnotesize Particle ratios for 200 GeV central Au+Au
            collisions at RHIC
            \cite{bearden,chujo,wosiek,lee,fuqiang,geno,debsankar}.
            The error bars depict the statistical errors, the brackets
            the systematic uncertainties.  In the upper row the
            vertical solid lines indicate the RHIC average, the dashed
            lines the error on the mean. Open circles are used for
            data points which were extracted from figures. Results are
            preliminary, except BRAHMS and PHOBOS data in upper row.}
        \label{fig:ratios}
        \vspace{-8mm}
    \end{center}
\end{figure}
The assumption of a locally thermalized source in chemical equilibrium
can be tested by using statistical thermal models to describe the
ratios of various emitted particles. This yields a baryon chemical
potential $\mu_B$, a strangeness saturation factor $\gamma_s$, and the
temperature $T_{\mathrm{ch}}$ at chemical freeze-out.  Because of the
absence of any dynamic assumptions many details can never be fully
absorbed by these models.  Discrepancies between model and data up to
30\% should be considered inside the systematic uncertainty of the
thermal model approach \cite{uli}.  So far these models are remarkably
successful in describing particle ratios at SPS
\cite{pbm-sps,becattiniSPS} and now also at RHIC
\cite{pbm-rhic,becattiniRHIC}. This observation, together with the
large collective flow (radial and elliptic) measured at RHIC, is
generally considered a strong hint that chemical equilibrium is indeed
reached.  The wide reaching implications of thermal statistical models
and the models themselves were the subject of a dedicated podium
discussion at the conference. I'm not going to summarize the issues
brought up in this discussion but rather will concentrate on the
recent experimental results on particle ratios and their implications
in the framework of the models.

Figure \ref{fig:ratios} shows a compilation of most results on
particle ratios at mid-rapidity for $\sqrtsNN = 200$ GeV presented at
the conference \cite{bearden,chujo,wosiek,lee,fuqiang,geno,debsankar}.
Given the good agreement among the experiments for the identical
particle ratios and the high level of quality of the data, it is
tempting to calculate the RHIC averages for these results. Adding the
statistical errors and the systematical uncertainties in quadrature to
derive the weights one obtains: $\pi^-/\pi^+ = 1.02 \pm 0.02$,
$K^-/K^+ = 0.95 \pm 0.03$, and $\overline{p}/p = 0.75 \pm 0.03$.  All
experiments reported to observe no \pt\ or centrality dependence of
these ratios for $\pt < 3$ \gevc, confirming earlier results at 130
GeV.  STAR reports a decrease in $\overline{p}/p$ at high \pt; for
discussion on this topic and also on the significant increase of the
$p/\pi$ ratios from low to medium \pt\ see \cite{kunde,Peitzmann}.

The analyses on non-identical particle ratios are not as complete and
in most cases the systematic uncertainties are still under evaluation.
While the net-baryon chemical potential at chemical freeze-out is
essentially determined by the baryon to antibaryon ratios
($\overline{p}/p,\overline{\Lambda}/\Lambda$ etc.), the non-identical
particle ratios are the ``thermometer'' of the thermal statistical
models. Because of the lack of sufficient constraints it would be
therefore premature to invoke the thermal model on the 200 GeV data at
this point. It is, however, instructive to compare the currently
available ratios with predictions made in \cite{pbm-rhic}. Here the
authors used a phenomenological parametrization of $\mu_B$, obtained
from thermal parameters derived from the statistical model at lower
energies in conjunction with the assumption of constant energy per
particle (see above) to extrapolate to 200 GeV.

\begin{figure}[t]
    \begin{center}
        \includegraphics[width=0.95\textwidth]{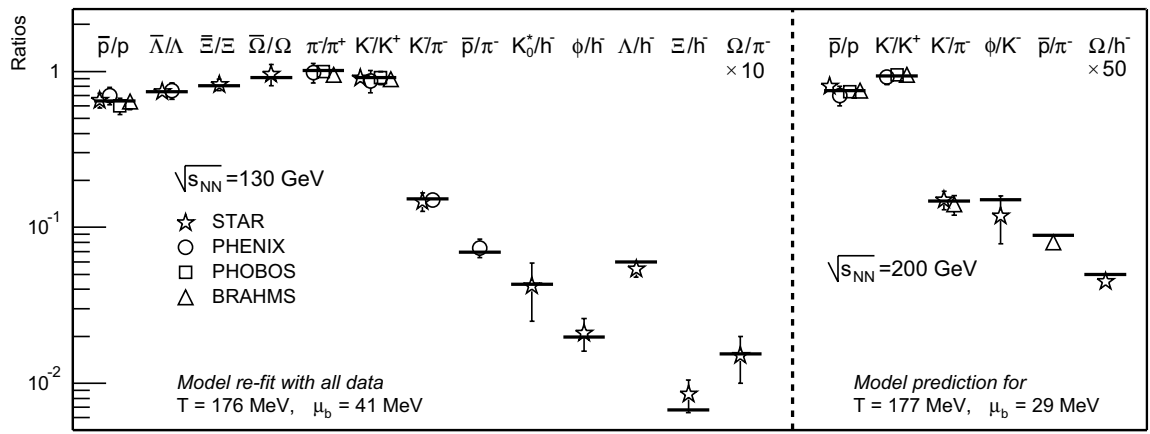}
        \vspace{-5mm}
        \caption{\footnotesize Left panel: comparison between RHIC
            experimental particle ratios for $\sqrtsNN = 130$ GeV and
            statistical model calculations with $T_{\mathrm{ch}} =
            176$ MeV and $\mu_B = 41$ MeV (from \cite{pbm-rhic,dan}).
            Right panel: comparison between RHIC ratios at $\sqrtsNN =
            200$ GeV and prediction discussed in the text (also
            \cite{pbm-rhic,dan}).}
        \label{fig:statModel}
    \end{center}
    \vspace{-1cm}
\end{figure}
This comparison, together with the statistical model fit for 130 GeV,
updated with the latest values presented at this conference is shown
in Fig.~\ref{fig:statModel} \cite{dan}. The predictions match well
with the current results and indicate no significant change in
$T_{\mathrm{ch}}$ but a drop in $\mu_B$ from $\sim 41$ MeV at
$\sqrtsNN = 130$ GeV to 29 MeV at 200 GeV. The latter value is also in
agreement with calculations made by various authors at the conference
using different approaches \cite{bearden,wosiek,chujo}. The chemical
freeze-out temperature is naturally limited by the confinement
phase-transition temperature assumed to be around 175 MeV, although
$T_{\mathrm{ch}}$ is actually not constrained in thermal model fits.

\begin{wrapfigure}{r}{0.5\textwidth}
    \vspace{-8mm}
    \includegraphics[width=0.5\textwidth]{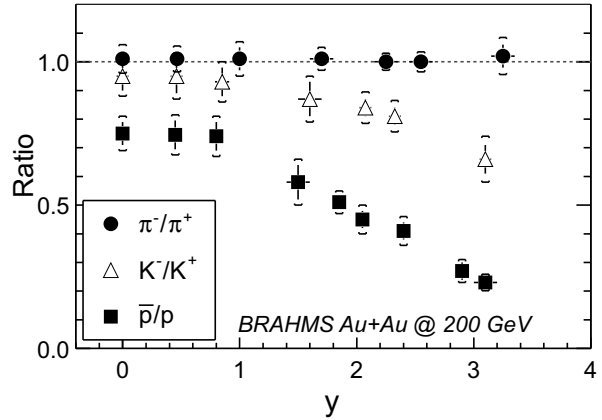}
    \vspace{-8mm}
        \caption{\footnotesize Antiparticle to particle ratios
            as a function of rapidity in 200 GeV Au+Au collisions
            (from \cite{bearden,lee}).}
        \label{fig:brahmsFig3}
    \vspace{-8mm}
\end{wrapfigure}
Another interesting result was reported by the BRAHMS collaboration
who presented a detailed study of identical particle ratios as a
function of rapidity for central events \cite{bearden,lee}.  As shown
in Fig.~\ref{fig:brahmsFig3} the $\pi^-/\pi^+$ ratio is consistent
with unity over the considered rapidity range while the $K^-/K^+$
ratios drops by $\sim 30$\% at $y = 3$ from its mid-rapidity value and
the $\overline{p}/p$ ratio by $\sim 70$\%. Interestingly, all ratios
remain constant for $|y|<1$, consistent with the assumption of boost
invariance around mid-rapidity (see below).  From those data the
authors derived a net-baryon chemical potential at $y = 3$ of $\mu_B \sim
130$ MeV within the framework of a statistical model, assuming that
the particle sources in the different $y$ regions are in local
chemical equilibrium and that strangeness is locally conserved.
However, to what extent a thermal interpretation at large forward
rapidities is justified is subject to further studies.

\section{Boost Invariance at RHIC?}

\begin{wrapfigure}{r}{0.5\textwidth}
    \vspace{-13mm}
    \includegraphics[width=0.5\textwidth]{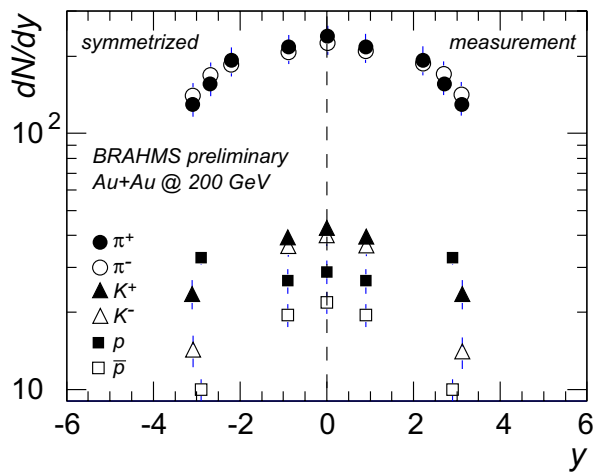}
    \vspace{-8mm}
    \caption{\footnotesize Rapidity distributions
        for $\pi^\pm, K^\pm, p$, and $\overline{p}$ in 200
        GeV Au+Au collisions for the top 10\% most central events
        (data taken from \cite{bearden,lee}).}
    \label{fig:richRap1}
    \vspace{-10mm}
\end{wrapfigure}
Many models assume directly or indirectly that the system created in
heavy ion collisions is boost invariant, \textit{i.e.}~invariant under
Lorentz transformations in the beam direction. Commonly, but not
correctly, one considers a system boost invariant within a given
rapidity interval if the rapidity distribution $dN/dy$ is constant
within that range.  Strictly speaking, however, it requires all
Lorentz invariant observables to remain constant.
The pseudorapidity distribution of charged particles at RHIC shows a
plateau extending over almost 3 units \cite{phobosRap}.  $dN/d\eta$
distributions, however, can not be used to draw any conclusions on
boost invariance because the Jacobian $\partial y/\partial\eta(\pt,
\eta)$ that tends to flatten otherwise peaked rapidity distributions.
The BRAHMS collaboration presented at this conference the rapidity
distribution of pions, kaons, and protons and their antiparticles over
6 units of rapidity ($|y|<3$) for Au+Au collisions at $\sqrtsNN = 200$
GeV \cite{bearden}.

\begin{wrapfigure}{l}{0.5\textwidth}
    \vspace{-0mm}
    \includegraphics[width=0.48\textwidth]{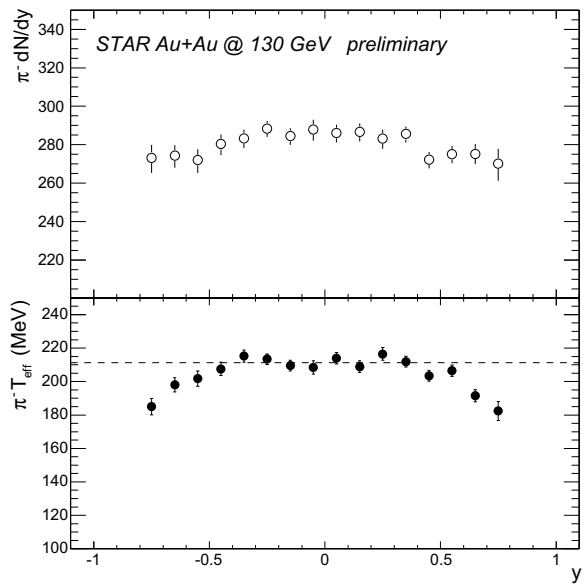}
    \vspace{-3mm}
    \caption{\footnotesize Upper
        panel: $\pi^-$ rapidity distribution in 130 GeV Au+Au
        collisions for top 5\% most central events. Lower panel:
        inverse slope obtained from Bose-Einstein fits to the \pt\ 
        spectra at different rapidities (from
        \cite{calderon,gene}).}
    \label{fig:richRap2}
    \vspace{-8mm}
\end{wrapfigure}
As shown in the compilation of all six distributions in
Fig.~\ref{fig:richRap1} and given the statistical errors and
systematic uncertainties a plateau is at most limited to $\pm 1$ unit
in rapidity. The upper panel in Fig.~\ref{fig:richRap2} shows the pion
rapidity distribution in this very region as measured by STAR with
higher granularity \cite{gene}.  A study of the slope of the
corresponding \pt-distribution, depicted in the lower panel, shows
that the slope starts to decrease from its mid-rapidity value at
around $\pm \eta \sim 0.5$ thus limiting the region of boost
invariance to $|\eta|<0.5$.  This is also confirmed for the case of
the protons by BRAHMS where no significant change in slope is observed
between between $y=0$ and 0.9 \cite{lee}.  In this context it is also
important to recall Fig.~\ref{fig:brahmsFig3} to verify that the
particle to antiparticle ratios of the most abundant particles ($\pi,
K$, and $p$) remain constant within half a unit around mid-rapidity.
Recent studies of elliptic flow also indicate a continous drop in
$v_2$ for $|\eta| \gtrsim 1$ that is, at least in parts, related to
the strong dependence of $v_2$ on $\pt(\eta)$ \cite{manly}.  We
conclude that the currently available RHIC data suggests a rather
small truely boost-invariant region of at most $|\eta|<0.5$.

\section{Inclusive Transverse Momentum Spectra and Radial Flow}
\label{sec:spectra}

Inclusive hadron spectra were intensively discussed at this conference
in the context of the observed suppression of high-\pt\ yields for
central collisions when compared to either peripheral collisions or
$pp$ (\ppbar) reference data \cite{Peitzmann}. These studies focus
naturally only on a tiny fraction of all produced particles. The
majority of particles emitted from the system are soft. At RHIC 99.9\%
of all charged particles have momenta below 2 \gevc, far outside the
range of perturbative QCD.

Transverse momentum spectra of identified particles reflect the system
at kinetic freeze-out and allow us to extract information from the
latest stage of the evolution when the system was still thermally
coupled and governed by elastic interactions among its constituents.
The measured inverse slope parameter is determined by two components:
the actual temperature at the freeze-out and the transverse flow
component. In simple terms this can be approximated as $T =
T_{\mathrm{fo}} + m \langle \beta_T \rangle^2$ where $\beta_T$ is the
transverse flow velocity.  This ansatz, however, has the disadvantage
that attempts to extract $T_{\mathrm{fo}}$ and $\beta_T$ are strongly
dependent on the range in which the slopes were determined. Of even
greater concern is the assumption of a fixed flow velocity which
oversimplifies the problem considerably.  To overcome these problems
and to avoid the complexity of adjusting the initial energy density
and the equation of state in a full hydrodynamical model calculation
many studies now use the so-called 'blastwave' parametrization
\cite{schnederman}. Here, the invariant cross-section is fit to
\begin{eqnarray}
    \frac{dN}{m_T dm_T} \propto \int^R_0 r dr m_T
    K_1(\frac{m_T \cosh \rho}{T_{\mathrm{fo}}})
        I_0(\frac{p_T \sinh \rho}{T_{\mathrm{fo}}})
\end{eqnarray}
where $\rho = \tanh^{-1} \beta_T$ is the transverse rapidity and
$\beta_T = \beta_s (r/R_{max})^n$ depends on the chosen flow profile
and the flow at the surface $\beta_s$. There is no commonly accepted
flow profile and $n$ varies in the different analysis between 0.5 and
2. It is important to keep in mind that $T_{\mathrm{fo}}$ and
$\beta_T$ are correlated. Increasing $T_{\mathrm{fo}}$ or $\beta_T$
has to some degree a similar effect on the spectral shape. This
problem can be overcome by applying a blastwave motivated
parametrization also to HBT radii, $R(k_T)$, since here
$T_{\mathrm{fo}}$ and $\beta_T$ are anti-correlated helping to further
constrain $T_{\mathrm{fo}}$ and thus $\beta_T$.

\begin{wrapfigure}{r}{0.60\textwidth}
    \vspace{-10mm}
    \includegraphics[width=0.60\textwidth]{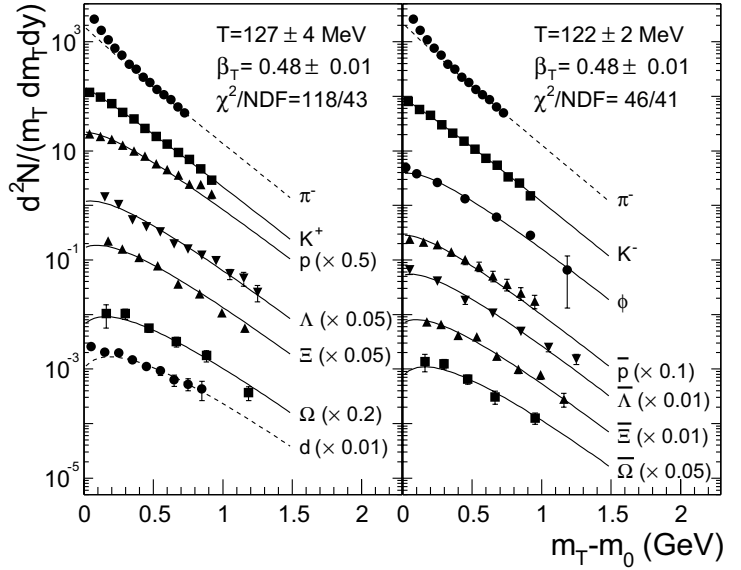}
    \vspace{-8mm}
    \caption{\footnotesize Transverse mass spectra in central 158 AGeV
        Pb+Pb collisions measured by NA49. The curves
        show the result of blastwave fits with a constant
        flow profile, \textit{i.e.} $n=0$ (from \cite{marco}).}
    \label{fig:MarcoPage11}
    \vspace{-7mm}
\end{wrapfigure}
Van Leeuwen presented an impressive compilation of NA49 transverse
mass spectra in 40, 60, and 158 AGeV collisions \cite{marco}. Figure
\ref{fig:MarcoPage11} depicts the results from Pb+Pb collisions at 158
AGeV where all spectra were fit to a blastwave parameterization.  The
obtained freeze-out parameters are $T_{\mathrm{fo}} = 122-127$ MeV and
$\langle \beta_T \rangle = 0.48 \pm 0.01$ in agreement with previous
studies.  Blastwave fits to RHIC data give a slightly smaller
freeze-out temperature $T_{\mathrm{fo}} \sim 110$ MeV but a higher
flow value $\langle \beta_T \rangle = 0.55 - 0.6$ due to the higher
pressure in the system \cite{gene,jane}.  The surprising finding in
the NA49 analysis, however, is the fact that not only are the $\pi, K,
p$, and $\Lambda$ spectra well described by the fit but so are the
spectra of the multi-strange baryons $\Xi$ and $\Omega$.  So far
multi-strange baryons were assumed to show less or no flow, due to a
possibly very small elastic cross-section.  This was supported by
simple fits to the $\Omega$ spectra which yield $T_{\mathrm{fo}}$
values close to the chemical freeze-out temperature.  It has been
speculated that if the elastic cross-sections for the $\Omega$ were
indeed very small, a non-zero flow component could be interpreted as a
signature for partonic flow. The situation at RHIC is still ambiguous
since the current $\Omega$ $m_T$-spectra lack statistics and within
the errors can be fit by either assuming $T_{\mathrm{fo}} \sim 170$
and $\langle \beta_T \rangle \sim 0.4$ or $T_{\mathrm{fo}} = 130$ and
$\langle \beta_T \rangle \sim 0.5$ \cite{suire}. One can, however,
already exclude the cases of no flow and, unlike at SPS, it cannot be
described by parameters obtained from fits to pions, kaons, and
protons alone. To what extent a combined fit of all spectra, similar
to that presented for the SPS, yields better results still has to be
seen.  Higher precision data is certainly needed here.

\begin{figure}[tbh]
    \begin{center}
        \includegraphics[width=0.9\textwidth]{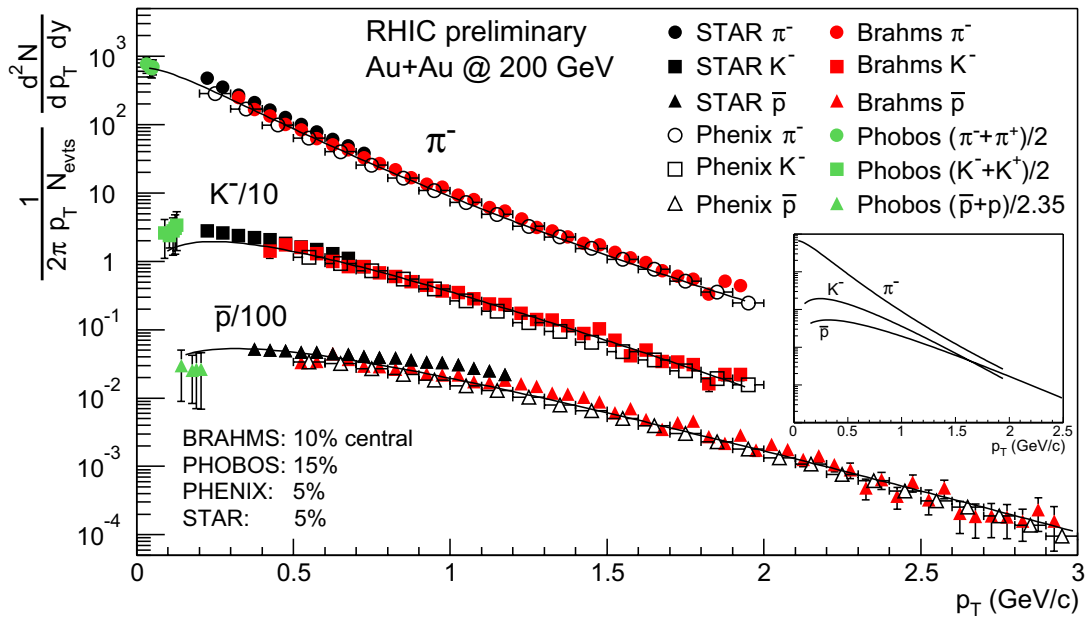}
        \vspace{-10mm}
        \caption{\footnotesize Compilation of preliminary transverse
            momentum spectra of $\pi^-, K^-,$ and $\overline{p}$ for
            200 GeV central Au+Au collisions from all RHIC
            experiments.  Except for the PHOBOS data all spectra are
            not feed-down corrected. The curves are fits to PHOBOS,
            BRAHMS, and PHENIX data. For $\pi^-$ a power law in $m_T$
            was used; Maxwell-Boltzmann distributions for $K^-$ and
            $\overline{p}$.}
        \label{fig:ptspectra}
        \vspace{-7mm}
    \end{center}
\end{figure}
Figure \ref{fig:ptspectra} shows a compilation of \pt\ spectra for
$\pi^-, K^-$ and $\overline{p}$ from all four RHIC experiments at
$\sqrtsNN = 200$ GeV. For the first time the PHOBOS collaboration
presented data at very low \pt, down to 30 MeV for pions, thus extending
our knowledge into a range which is not accessible to the other three
experiments. The low-\pt\ region is very sensitive to dynamic effects
and will help to constraint models even further.
Except for PHOBOS all spectra shown in the figure are not feed-down
corrected.  It is important to note, that the RHIC experiments are not
equally sensitive to feed-down, e.g.~BRAHMS because of its small
aperture is less affected than a large acceptance detector such as STAR.
Note, also the slightly different centrality selections. The
difference in yields between 15\% and 5\% centrality is approximately
15\%.  One of the problems in determining the absolute yield of
particles is the need to extrapolate to $\pt = 0$. At RHIC this results
in considerable uncertainties in the yields of typically $\sim 5-10$\%
for experiments limited to $\pt > 200$ \mevc\ or above. The new
low-\pt\ data now allows us to significantly reduce this source of
uncertainty.  It is interesting to note that pion spectra actually
do not follow a power-law in \pt\ down to low \pt\ as indicated by
earlier measurements, but can be very well described by a power-law in
$m_T$: $\propto\ A\,(1+m_T/m_0)^{-n}$ as indicated by the fit shown in
Fig.~\ref{fig:ptspectra}.

Radial flow affects the spectra of heavier particles considerably more
than light particles. One of the consequences of strong radial flow at
RHIC is that the yields of pions, kaons, and protons essentially
become equal at around $\pt = 2$ \gevc\ as depicted in the smaller
panel in Fig.~\ref{fig:ptspectra}. At large \pt\ the baryon to meson
ratios should drop again as flow should affect high-\pt\ particles to
a much lesser degree and the ratios should approach values predicted
by pQCD.  Other studies follow a different approach, invoking novel
baryon dynamics attributed to gluonic baryon junctions that predict
the baryon-enhancement only in a finite moderate-\pt\ window
\cite{vitev}.  However, the 'turn-over' point, predicted in these
models to be around 3 \gevc, has not yet been clearly identified
\cite{chujo}.

\section{Summary}
\label{sec:summary}

The recent results highlighted at the conference and in this review
show the enormous efforts of the community to study 'bulk' matter
governed by the non-perturbative regime of QCD. The new RHIC data
recorded at $\sqrtsNN = 200$ GeV seem, without exceptions, to confirm
the picture that evolved from the first data, at 130 GeV, reported at
Quark Matter 2001. This picture cannot be understood in terms of
global variables, ratios, and spectra alone but only by combining the
information we get from studies of HBT, elliptic flow, and high-\pt,
to name only a few.

Although we observe quantitative differences between SPS and RHIC in
many parameters, there appears not to be any striking qualitative
difference between these two energy regimes with the very prominent
exception of the onset of hard scattering processes at RHIC.

The study and interpretation of soft hadron spectra, particle ratios,
and yields is dominated by the success of the thermal statistical
models. Although critical issues concerning these models need to be
resolved, it remains a fact that they describe the ratios and yields
remarkably well over a wide range of energies. They indicate that in
the energy range from a few GeV up to RHIC energies the observed hadrons 
originate from a system in chemical equilibrium along a unified
freeze-out curve. This curve provides the relation between the
temperature and the baryon chemical potential. At RHIC we are
approaching an almost net-baryon free system with $\mu_B \sim 25$ MeV
where, for the first time, more baryons are produced at mid-rapidity
than transported from beam-rapidity. The chemical freeze-out
temperatures at SPS ($T_{\mathrm{ch}} \sim 165$ MeV) and at RHIC
($T_{\mathrm{ch}} \sim 175$ MeV) extracted from these models appear to
be close to the critical temperature. This implies that chemical
equilibrium is not caused by kinetic equilibration through hadronic
rescattering but indicates that hadron formation proceeds by
statistical hadronization from a prehadronic state.

Detailed analysis of the hadron spectra shows that the system expands
collectively under strong internal pressure. Radial flow at RHIC
appears to be slightly higher than at SPS, the kinetic freeze-out
temperatures are very close to each other, possibly somewhat lower at
RHIC ($T_{\mathrm{fo}} \sim 110$ MeV) than at SPS ($T_{\mathrm{fo}}
\sim 120-130$ MeV).  Studies of resonances \cite{patricia} and
correlations \cite{lanny} show that the time scale for emission is
very short (2-3 fm/$c$) while the overall lifetime of the system
appears to be on the order of 10 fm/$c$. This implies that after
hadronization the hadron abundances freeze out more or less
immediately.

We could declare proof of the quark-gluon plasma on the basis of
indirect evidence, but the fact that a new phase exists is almost
trivial compared to characterizing its features.  The interpretation
of bulk properties in heavy ion collisions was, and still is, complex.
We have evidence for thermalization of hot and dense matter and we
have indications of unusual behavior in rare, high momentum probes.
The level of collectivity is surprising but the timescales are
puzzling. We observe matter that is surely not a simple collection of
elementary particles, and we have the tools to study it.
 
Let me finally thank all the speakers at the conference who have
provided me with their results. I am grateful to many colleagues for
critical discussions and useful support, in particular to M.~Baker,
R.~Bellwied, M.~Calderon, H.~Caines, J.~Harris, B.~Hippolyte, F.~Laue,
D.~Magestro, K.~Redlich, C.~Suire, and Z.~Xu.

\end{document}